\documentclass[
 reprint,
superscriptaddress,
amsmath,
amssymb,
aps,
reprint,
superscriptaddress
]{revtex4-1}

\usepackage[T1]{fontenc}
\usepackage{graphicx}
\usepackage{dcolumn}
\usepackage{bm}
\usepackage{braket}
\usepackage{mathptmx}
\usepackage{color}
\usepackage{xcolor}
\usepackage[normalem]{ulem}

\begin{document}
\title{Generating Gottesman-Kitaev-Preskill qubit using cross-Kerr interaction\\between squeezed light and Fock states in optics}
\author{Kosuke Fukui}
\affiliation{%
Department of Applied Physics, School of Engineering, The University of Tokyo, 7-3-1 Hongo, Bunkyo-ku, Tokyo 113-8656, Japan}
\author{Mamoru Endo}
\affiliation{%
Department of Applied Physics, School of Engineering, The University of Tokyo, 7-3-1 Hongo, Bunkyo-ku, Tokyo 113-8656, Japan}
\author{Warit Asavanant}
\affiliation{%
Department of Applied Physics, School of Engineering, The University of Tokyo, 7-3-1 Hongo, Bunkyo-ku, Tokyo 113-8656, Japan}
\author{Atsushi Sakaguchi}
\affiliation{%
Optical Quantum Computing Research Team, RIKEN Center for Quantum Computing, 2-1 Hirosawa, Wako, Saitama 351-0198, Japan}
\author{Jun-ichi Yoshikawa}
\affiliation{%
Department of Applied Physics, School of Engineering, The University of Tokyo, 7-3-1 Hongo, Bunkyo-ku, Tokyo 113-8656, Japan}
\author{Akira Furusawa} 
\affiliation{%
Department of Applied Physics, School of Engineering, The University of Tokyo, 7-3-1 Hongo, Bunkyo-ku, Tokyo 113-8656, Japan}
\affiliation{%
Optical Quantum Computing Research Team, RIKEN Center for Quantum Computing, 2-1 Hirosawa, Wako, Saitama 351-0198, Japan}
\begin{abstract}
Gottesman-Kitaev-Preskill (GKP) qubit is a promising ingredient for fault-tolerant quantum computation (FTQC) in optical continuous variables due to its advantage of noise tolerance and scalability. However, one of the main problems in the preparation of the optical GKP qubit is the difficulty in obtaining the nonlinearity. 
Cross-Kerr interaction is one of the promising candidates for this nonlinearity.
There is no existing scheme to use the cross-Kerr interaction to generate the optical GKP qubit for FTQC.
In this work, we propose a generation method of the GKP qubit by using a cross-Kerr interaction between a squeezed light and a superposition of Fock states. 
We numerically show that the GKP qubit with the 10 dB can be generated with a mean fidelities of 99.99 and 99.9~\% at the success probabilities of 2.7 and 4.8~\%, respectively. 
Therefore, our method has potential method to generate the optical GKP qubit with a quality required for FTQC when we 
obtain the sufficient technologies for the preparation of ancillary Fock states and a cross-Kerr interaction.
\end{abstract}

\maketitle

\section{Introduction}
Quantum information processing with continuous variables (CVs)~\cite{lloyd1999quantum,bartlett2002efficient} has been receiving attentions for few decades.
In a CV system, the bosonic code for encoding quantum information in CVs is essential to remove errors during quantum information processing.
A variety of bosonic codes have been developed, e.g., the cat code~\cite{chuang1997bosonic,cochrane1999macroscopically,albert2019pair,niset2008experimentally} and the binomial code~\cite{michael2016new}. 
Among bosonic codes, the Gottesman-Kitaev-Preskill (GKP) qubit~\cite{gottesman2001encoding} is a promising way to encode a qubit in CVs, where qubit is encoded in the continuous Hilbert space of harmonic oscillator's position and momentum variables. 
In particular, the GKP qubit has advantages of an error tolerance and scalability towards large-scale quantum computation (QC) with CVs: 
(1) The GKP qubit is designed to protect against small displacement noise, and can achieve the hashing bound of the additive Gaussian noise with a suitable quantum error correcting code~\cite{fukui2017analog,fukui2018high}. 
Furthermore, it protects against other types of noise, including noise from finite squeezing during measurement-based QC~\cite{menicucci2014fault} and photon loss~\cite{albert2018performance}. 
(2) The GKP qubit inherits the advantage of squeezed vacuum states on optical implementation; they can be entangled by only beam-splitter coupling~\cite{note1}.
This feature allows us to entangle the GKP qubit with large-scale CV cluster state by using beam-splitter coupling.
Recently, a large-scale, two-dimensional CV cluster state composed of squeezed vacuum states, has been realized experimentally in an optical setup~\cite{asavanant2019generation,larsen2019deterministic}.
In addition, the single-and two-mode gates on a large-scale cluster state has also been demonstrated~\cite{Asavanant2021,larsen2021deterministic}.
From these advantages, the GKP qubit is an indispensable resource for fault-tolerant quantum computation (FTQC) with CVs~\cite{menicucci2014fault,baragiola2019all,pantaleoni2020modular,
walshe2020continuous,pantaleoni2021subsystem,grimsmo2021quantum,fukui2018tracking,vuillot2019quantum,noh2020fault,noh2020encoding,noh2021low,fukui2021efficient,seshadreesan2021coherent}.
Furthermore, the GKP qubit is a promising element to a variety of quantum information processing such as long-distance quantum communication~\cite{fukui2021all,rozpkedek2021quantum}.

Experimentally, the GKP qubit has been generated recently in an ion trap~\cite{fluhmann2019encoding} and superconducting circuit quantum electrodynamics~\cite{campagne2020quantum}, where the squeezing level of generated GKP qubits is close to 10 dB, 
the required squeezing level for FTQC with CVs~\cite{fukui2018high,fukui2019high,noh2020fault,yamasaki2020polylog,
bourassa2021blueprint,tzitrin2021fault}.
On the other hand, a large-scale CV cluster state has so far not been demonstrated in such physical setups. 
Thus, generations of the GKP qubit may not be directly translated into large-scale QC except for an optical setup, since the advantage of the GKP qubit for scalability in CVs has not been employed as far as a framework of measurement-based QC with CVs~\cite{note4}.
In an optical setup, 
various generation methods of GKP state are being developed
~\cite{pirandola2004constructing, pirandola2006continuous, pirandola2006generating, motes2017encoding,eaton2019non,su2019conversion,
arrazola2019machine,tzitrin2020progress,lin2020encoding,hastrup2021generation,fukui2021efficient2}.
Unfortunately, however, optical generation of the GKP qubit has remained unsuccessful due to the difficulties in obtaining the required nonlinearity

The Kerr type effect---a nonlinear effect where the refractive index onf a material changes when an electrical field is applied---has been widely studied as one of the methods to obtain a nonlinearity in an optical setup.
For classical information processing, self-and cross-Kerr effects have been widely studied~\cite{dudley2009ten,dudley2009modulation,kupchak2019terahertz,endo2021coherent}, where self-and cross-Kerr effects change the refractive index depending on the incoming field itself and another field, respectively.
In this work, we focus on the use of a cross-Kerr effect for the GKP qubit generation.
The cross-Kerr effect has been studied extensively for quantum information processing~\cite{jeong2004generation,van2006hybrid,jin2007generating,glancy2008methods,
lin2009quantum,feizpour2011amplifying,venkataraman2013phase,schmid2017verifying,
yanagimoto2020engineering}, such as quantum computation~\cite{nemoto2004nearly,munro2005weak,spiller2006quantum,jeong2006quantum,
shapiro2007continuous}, and the proposal to apply the cross-Kerr effect to generate the optical cat code, which is a typical non-Gaussian state for quantum information processing~\cite{gerry1999generation,jeong2005using,glancy2008methods,he2009scheme}. 

The scheme for the GKP qubit generation with the cross-Kerr interaction has been proposed in Ref.~\cite{pirandola2004constructing}. 
In this scheme, two coherent states interact with each other via the cross-Kerr interaction, and the GKP qubit is generated after the measurement of one of the coherent states.
One merit of this scheme is that the effective
interaction could be large by using a large amplitude of the coherent state, even if the cross-Kerr interaction per photon is small.
On the other hand, it is an open problem whether the state generated by Ref.~\cite{pirandola2004constructing} is appropriate for FTQC with the GKP qubit or not
due to the two features: the first one is that the generated state has different codewords between position and momentum quadratures. Since the two-qubit gate on the GKP qubits interacts with the quadratures with each other, it is more suitable to use the same codewords between position and momentum quadratures.
This feature requires the additional squeezing operation to match the codewords between both quadratures, where the squeezing operation in an optical setup~\cite{yoshikawa2007demonstration, miyata2014experimental} introduces a noise derived from a finite squeezing. The second one is that the probability of misidentifying the bit value of the generated state is $\sim 1\%$.
The threshold of the squeezing level for FTQC with the GKP qubit has been known to be around 10 dB at most~\cite{fukui2018high,fukui2019high,yamasaki2020polylog}, which corresponds to the probability of the misidentifying the bit value, $\sim 0.01\%$.
Although it is not known whether the probability of misidentifying the bit value of the state in Ref.~\cite{pirandola2004constructing} can be improved, 
it is unclear currently that the generated state could be FTQC with the GKP qubit~\cite{note6}.

In this paper, we propose the method for generating the GKP qubit via a cross-Kerr interaction between a squeezed light and the ancilla state of a superposition of Fock states, instead of the interaction between two coherent states in the conventional method.
However, there is another difficulty due to the rotation of squeezed light in phase space, when replacing the coherent state with a squeezed light, as described in Sec.~III.
Our scheme circumvents this problem by using positive and negative amplitudes for a cross-Kerr interaction, allowing us to apply the squeezed light to the preparation of the GKP qubit via a cross-Kerr interaction. In addition, the squeezing level of the generated GKP qubit is equal to the initial squeezing level of the input state.
In the numerical calculation, we show that our method can prepare GKP qubit with a high squeezing level and high fidelity. Thus, our method has the potential towards FTQC with the GKP qubit if we can prepare the input squeezed vacuum state with a sufficient squeezing and the ancilla state of a superposition of Fock states.

The rest of the paper is organized as follows. In Sec.~\ref{Sec2}, we review the GKP qubit, the cross-Kerr effect, and the conventional scheme of the GKP qubit generation by using a cross-Kerr interaction.
In Sec.~\ref{Sec3}, we describe our scheme that uses a cross-Kerr interaction between a squeezed light and a superposition of Fock states.
In Sec.~\ref{Sec4}, we numerically investigate our scheme, showing numerical calculations of the fidelity between the generated state and the target GKP qubit, and the success probability of our scheme.
Sec.~\ref{Sec5} is devoted to discussion and conclusion. 

\section{Background}\label{Sec2}
In this section, we review the GKP qubit, a simple example to apply a cross-Kerr interaction to a quantum information processing, and the conventional scheme to generate the GKP qubit by using a cross-Kerr interaction.

\subsection{GKP qubit}
In this article, we work in units where $\hbar=1$ and the vacuum variances are $\langle\hat{q}^2\rangle_\text{vac}=\langle\hat{p}^2\rangle_\text{vac}=1/2$ for the position quadrature $\hat{q}$ and momentum quadratures $\hat{p}$. The logical 0 and 1 states of the GKP qubit $\ket {\widetilde{0}}$ and $\ket {\widetilde{1}}$ are composed of a series of Gaussian peaks of width $\Delta$ contained in a larger Gaussian envelope of width 1/$\kappa$, and each peak of logical 0 (1) state is separated by $2\sqrt{\pi}$ each other. In the position basis, the logical states $\ket {\widetilde{0}}$ and $\ket {\widetilde{1}}$ are given by
\begin{eqnarray}
\ket {\widetilde{0}} &\propto &  \sum_{t=- \infty}^{\infty} \int e^{-\frac{(2t\sqrt{\pi})^2}{2(1/\kappa^2)}}{e}^{-\frac{(q-2t\sqrt{\pi})^2}{2\Delta^2}}\ket{q}  dq,    \\ 
\ket {\widetilde{1}} &\propto &  \sum_{t=- \infty}^{\infty} \int e^{-\frac{[(2t+1)\sqrt{\pi}]^2}{2(1/\kappa^2)}}  
{e}^{-\frac{(q-(2t+1)\sqrt{\pi})^2}{2\Delta^2}}\ket{q}  dq .     \label{gkp}
\end{eqnarray}

Although these states become the perfect GKP qubits with Dirac-comb wavefunctions in case of infinite squeezing ($\Delta \rightarrow 0$, $\kappa \rightarrow 0$)~\cite{gottesman2001encoding}, they are not orthogonal in the finite squeezing regime
and there is a non-zero probability of misidentifying $\ket {\widetilde{0}}$ with $\ket {\widetilde{1}}$, and vice versa. We choose $\kappa$ and $\Delta$ so that the variance of each peak in the position and momentum observables is equal to $\sigma^{2}$, i.e., $\Delta^{2} = \kappa^{2} = 2\sigma^{2}$. 
The squeezing level $s$ is equal to $s=-10{\rm log}_{10}2\sigma^2.$

\subsection{Cross-Kerr interaction for the controlled-NOT gate}
As an example of a cross-Kerr interaction for quantum information processing, we describe the controlled-NOT gate between two photons~\cite{nemoto2004nearly,munro2005weak}, which is a typical way to use this interaction for quantum information processing.
We consider the interaction with the cross-Kerr interaction between two modes a and b. The Hamiltonian for a cross-Kerr interaction, $\hat{H}_{\rm CK}$, is described by
\begin{equation}
\hat{H}_{\rm CK}=\hbar \chi \hat{a}^{\dag}\hat{a}\hat{b}^{\dag}\hat{b},
\end{equation}
where $\chi$ is the strength of the cross-Kerr nonlinearity, $\hat{a}(\hat{b})$ and $\hat{a}^\dag(\hat{b}^\dag)$ are the annihilation and creation operators for mode a (b), respectively. The annihilation operators are given by $\hat{a}(\hat{b}) = (\hat{q}_{\rm a(b)}+i\hat{p}_{\rm a(b)})/\sqrt{2}$.
We then see the interaction between a superposition of Fock states, $\ket{\phi}_{\rm a}=(\ket{0}_{\rm a}+\ket{1}_{\rm a})/\sqrt{2}$, and a coherent state, $\ket{\alpha}_{\rm b}$, where $\alpha$, $\ket{0}$, and $\ket{1}$ are the amplitude of the coherent state, a vacuum state, and a single photon, respectively.
After the interaction, the two states are entangled as 
\begin{eqnarray}
{e}^{-i\frac{\hat{H}_{\rm CK}}{\hbar}t}\ket{\phi}_{\rm a}\ket{\alpha}&=&^{-i\frac{\hat{H}_{\rm CK}}{\hbar}t}(\ket{0}_{\rm a}+\ket{1}_{\rm a})\ket{\alpha}_{\rm b} /\sqrt{2}\nonumber \\
&=& (\ket{0}_{\rm a} \ket{\alpha}_{\rm b}+\ket{1}_{\rm a} \ket{\alpha {e}^{i\theta}}_{\rm b}) /\sqrt{2},
\end{eqnarray}
where $\theta=\chi t$ with the interaction time $t$.
In addition to the above interaction between modes a and b, the coherent state b and the additional photon c interact with each other by the cross-Kerr interaction with the strength of the nonlinearity $-\chi$. After the interaction, the states are described as 
\begin{eqnarray}
e^{-i\frac{\hat{H}_{\rm CK}}{\hbar}t}&\ket{\phi}_{\rm c}& (\ket{0}_{\rm a} \ket{\alpha}_{\rm b}+\ket{1}_{\rm a} \ket{\alpha {e}^{i\theta}}_{\rm b})/{\sqrt{2}}\nonumber \\
=  \{(&\ket{0}_{\rm a}&\ket{0}_{\rm c}+ \ket{1}_{\rm a}\ket{1}_{\rm c} )\ket{\alpha}_{\rm b}\nonumber \\
+&\ket{1}_{\rm a}&\ket{0}_{\rm c} \ket{\alpha {e}^{i\theta}}_{\rm b}+\ket{0}_{\rm a}\ket{1}_{\rm c} \ket{\alpha {e}^{-i\theta}}_{\rm b}\}/2.
\end{eqnarray}
To disentangle the coherent state with photons, the homodyne measurement is performed on the coherent state.
If the measurement outcome $x$ is larger than $\alpha(1+{\rm cos}\theta)/2$, i.e., one distinguishes the coherent state as $\ket{\alpha}_{\rm b}$, the state is approximated by
$(\ket{0}_{\rm a}\ket{0}_{\rm c} +\ket{1}_{\rm a}\ket{1}_{\rm c} )/\sqrt{2}.$
On the other hand, if $x$ is smaller than $\alpha(1+{\rm cos}\theta)/2$, the state is approximated by
$({e}^{i \beta} \ket{0}_{\rm a}\ket{0}_{\rm c} +{e}^{-i \beta}\ket{1}_{\rm a}\ket{1}_{\rm c} )/\sqrt{2}, $
since the two states $\ket{\alpha {e}^{\pm i\theta}}$ are not distinguished, where $\beta=2\alpha {\rm sin} \theta (x-\alpha {\rm cos}\theta)$. Since the phase shift $\beta$ can be eliminated via a classical feed-forward operation~\cite{nemoto2004nearly,munro2005weak}, the states in modes a and c are entangled as
\begin{equation}
(\ket{0}_{\rm a}\ket{0}_{\rm c} +\ket{1}_{\rm a}\ket{1}_{\rm c} )/\sqrt{2}.
\end{equation}
The probability of misidentifying $\ket{\alpha}$ as $\ket{ \alpha {e}^{\pm i\theta}}$ is calculated by $\sim 10^{-4}$ when $\alpha \theta \sim \pi$ and $\alpha\sim 3\times 10^{5}$.
In Refs.~\cite{nemoto2004nearly,munro2005weak}, a weak nonlinearity is assumed to be $\theta \sim 10^{-5}$, which is achievable in the experiments~\cite{feizpour2015observation}.

\subsection{Conventional method for the GKP qubit generation}
\begin{figure}[t]
\centering\includegraphics[scale=0.7]{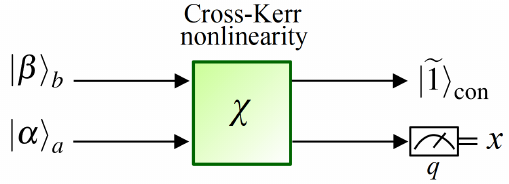}
\caption{A schematic diagram of the conventional method using cross-Kerr interaction~\cite{pirandola2004constructing}. $\ket{\alpha}_{\rm a}$ and $\ket{\beta}_{\rm b}$ with $\beta \gg \alpha$ are the input of coherent states are interacted with a Ker medium with a strength of the cross-Kerr nonlinearity, $\chi$. $\ket {\widetilde{1}}_{\rm con}$ is the generated state using the conventional method. $x$ is the outcome of the homodyne measurement in the $q$ quadrature. The generated state depends on the outcome $x$, where the generated state is close to the target GKP qubit $\ket {\widetilde{1}}$ when $x$ is close to zero.  } 
\label{fig1}
\end{figure}

In Ref.~\cite{pirandola2004constructing}, the cross-Kerr interaction between the two optical modes is used to generate the GKP qubit, where the two states in modes a and b are both prepared in coherent states as $\ket{\alpha}$ and $\ket{\beta}$ $( \alpha,~\beta \in  \mathbb{R})$ with $\beta \gg \alpha$, respectively.
Figure \ref{fig1} shows the schematic diagram for the conventional method of the GKP qubit generation, $\ket {\widetilde{1}}_{\rm con}$, which targets the logical 1 state of the GKP qubit, $\ket {\widetilde{1}}$.
The modes a and b interact with each other by a nonlinear medium, and mode a is measured by using the homodyne detector in the $q$ quadrature.
The two states after the interaction become
\begin{equation}
e^{-i\frac{\hat{H}_{\rm CK}}{\hbar}t}\ket{\alpha}_{\rm a}\ket{\beta}_{\rm b}={e}^{-\alpha^2/2}\sum_{n=0}^{\infty}\frac{\alpha^n}{\sqrt{n!}}\ket{n}_{\rm a}\ket{\beta {e}^{-in\theta}}_{\rm b}, \label{conv}
\end{equation}
where $\ket{n}_{\rm a}$ represents Fock states with photon number $n$ for mode $a$.
Assume that the cross-Kerr phase shift per photon, $\theta=\chi t$, is small, and $\ket{\beta {e}^{-in\theta}}_{\rm b}$ is approximated as $\ket{\beta -i n\beta\theta}_{\rm b}$ with $\chi t \alpha^2\ll 1$. 
To see the effect depending on $n$, 
we redefine $\ket{\beta -i n\beta\theta}_{\rm b}$ as $\ket{-i n\beta\theta}_{\rm b}$, respectively.
Then the right hand side of Eq.~(\ref{conv}) can be redefined as
\begin{equation}
{e}^{-\alpha^2/2}\sum_{n=0}^{\infty}\frac{\alpha^n}{\sqrt{n!}}\ket{n}_{\rm a}\ket{ -in\beta\theta}_{\rm b}, \label{conv2}
\end{equation}
which means that a large $\beta$ provides a sufficient interaction to realize a large displacement in the quadrature even if $\theta$ is small.

After the homodyne measurement of mode a, the generated state is described by the position and momentum wave functions, $\phi_q(q; \tau, \alpha, x)$ and $\phi_p(p; \tau, \alpha, x)$, as
\begin{eqnarray}
\phi_q(q; \tau, \alpha, x)\propto \sum_{n=0}^{\infty}\eta_n(\alpha, x) {\rm exp}(-q^2/2+i\pi n \tau q), \label{proq}\\
\phi_p(p; \tau, \alpha, x)\propto \sum_{n=0}^{\infty}\eta_n(\alpha, x) {\rm exp}[-(p-\pi n \tau)^2/2]\label{prop},
\end{eqnarray}
where $x$ is the outcome of the homodyne measurement, $\tau$ is the interaction time $\tau=-\sqrt{2}\beta \chi t/\pi$, and $\eta_n(\alpha, x)$ is defined as $\rho_n(\alpha,x){e}^{(\alpha^2+x^2)/2}$ with $\rho_n(\alpha,x)=\alpha^2 H_n(x)/(2^{n/2}n!)$ and the Hermite polynomials $H_n(x)$. 
In Ref.~\cite{pirandola2004constructing}, $\tau=2$ is assumed, and thus $\beta$ is needed to be $\beta\sim 4\times 10^{5}$ using $\chi t = \theta \sim10^{-5}$.

To implement FTQC with the GKP qubit, there are two features that the generated state may be not appropriate to implement fault-tolerant QC with the GKP qubit: The first one is the mismatch of the codewords between $q$ and $p$ quadratures. The second one is the probability of misidentifying the bit value for FTQC with the GKP qubits.
For the first problem, as we see Eqs.~(\ref{proq}) and (\ref{prop}), the intervals between codewords for the proposed method are different from those for the GKP qubit, $\sqrt{\pi}$. 
For example, the interval between codewords in the $q$ and $p$ quadrature for $\tau=\alpha=2$, becomes 1/4 and $4{\pi}$, respectively~\cite{pirandola2004constructing}. 
In general, this mismatch of the codewords makes it difficult to implement the two-qubit gate, e.g, the controlled-NOT gate or beam-splitter coupling, where the quadratures in $q$ and $p$ quadratures interact with each other. 
Thus, to implement the two-qubit gate, the additional squeezing operation to match the codewords will be required, where the squeezing operation in an optical setup introduces a noise derived from a finite squeezing~\cite{yoshikawa2007demonstration, miyata2014experimental}.
The second problem is regarding the probability of misidentifying the bit value.
The probability of misidentifying the bit value of the state generated by Ref.~\cite{pirandola2004constructing} is at least 1\%, which corresponds to that of the GKP qubit with a squeezing level~$\sim$6.2 dB. On the other hand, the threshold of the squeezing level is around 10 dB at least~\cite{fukui2018high,fukui2019high}, which corresponds to the probability of misidentifying the bit value~$\sim$0.01\%.
Since the error probability for the state generated by Ref.~\cite{pirandola2004constructing} is larger than that of the threshold value for FTQC with the GKP qubits, the state by Ref.~\cite{pirandola2004constructing} may be not sufficient to implement FTQC.
For the above two reasons, it has been unclear that the state generated by Ref.~\cite{pirandola2004constructing} could be FTQC with the GKP qubit.

\begin{figure*}[t]
\centering\includegraphics[scale=2.0]{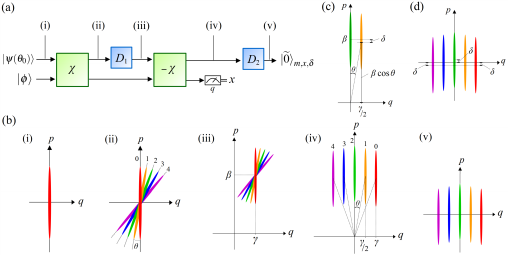}
\caption{A schematic diagram of the proposed method to generate. (a) (i) A superposition of Fock states, $\ket{\phi}$, and the squeezed light, $\ket{\psi(\theta_0)}$, are prepared. (ii) Two states are interacted via the cross-Kerr interaction with $\chi$. (iii) The displacement operation is performed on the squeezed light by $D_{1}=\beta+i\gamma$, where $\gamma=2m\sqrt{\pi}$ for the generation of $\ket {\widetilde{0}}_m$. (iv) Two states interact with each other by the cross-Kerr interaction with $-\chi$, which is an inverse amount of the first Kerr medium. (v) After the measurement of the superposition of Fock states, the displacement operation is performed on the squeezed light by $D_{2}=-\beta-m\delta/2$.
(b) A schematic description of the time evolution of the squeezed light in phase space, where (i)-(v) correspond to (a)(i)-(v), respectively.
We note that the Wigner representation is not used for the description of our method. 
Our aim here is to give an insight into how to translate the position of each squeezed light.
(c) The relation among $\beta$, $\gamma$, $\theta$, and $\delta$ in phase space. 
(d) The displacement error $\delta$ corresponds to $\beta(1-{\rm cos}\theta)$ with ${\rm cos}\theta=\sqrt{1-(\gamma/m\beta)^2}$. (b)-(d) show the case for $m=2$ as an example.
} 
\label{fig2}
\end{figure*}

\section{Proposed method}\label{Sec3}
In this work, we overcome the two problems of the conventional GKP qubit generation; the mismatch of the interval for codewords between both quadratures and the error probability of the generated GKP qubit.
To solve these problems, we use the squeezed light and a superposition of the Fock states with the appropriate coefficient weights as ancilla light for encoding the GKP code.
It may be natural that the squeezed light is expected to be applied to the conventional method, which employs the coherent light. 
However, it is difficult to employ the squeezed light because each light has a different rotation phase.
The key to evading this problem is to use the inverse of sign for the cross-Kerr interaction, as we will see later.

We here consider the generation of the GKP qubit with a finite number of peaks, e.g., the target GKP qubits, $\ket {\widetilde{0}}_m$ and $\ket {\widetilde{1}}_m$, are described as 
\begin{eqnarray}
\ket {\widetilde{0}}_m &\propto&   \sum_{t=-m}^{m} \int e^{-\frac{(2t\sqrt{\pi})^2}{2(1/\kappa^2)}}{e}^{-\frac{(s-2t\sqrt{\pi})^2}{2\Delta^2}}\ket{s}_q  ds, \label{gkpfinite0} \\
\ket {\widetilde{1}}_m &\propto &  \sum_{t=- m}^{m} \int e^{-\frac{[(2t+1)\sqrt{\pi}]^2}{2(1/\kappa^2)}}   
{e}^{-\frac{(s-(2t+1)\sqrt{\pi})^2}{2\Delta^2}}\ket{s}_q  ds, \label{gkpfinite1}
\end{eqnarray}
where $2m+1$ is the number of peaks, assuming m > 0 in this work~\cite{note5}.

In the following, we target the preparation of $\ket {\widetilde{0}}_m$.
Fig.~\ref{fig2} shows a schematic description of the proposed scheme consisting of a squeezed light, a superposition of Fock states, two Kerr mediums, and linear optics.
Our method consists of 5 steps, where each step corresponds to Fig.~\ref{fig2}(a)(i)-(v), respectively.
In Fig.~\ref{fig2}(b), the Wigner representation is not used for the description of our method, since we aim to give an insight into how to translate a position of each squeezed light.
In the first step, the squeezed light and a superposition of Fock states are prepared, as described in Fig.~\ref{fig2}(a)(i). A superposition of Fock states, referred to as the ancilla Fock state, is defined by 
\begin{equation}
\ket{\phi}=N\sum_{t=0}^{2m}c_{t}\ket{2t}, \label{ancilla}
\end{equation}
where $N$ is a normalization factor.
To obtain the envelope's distribution of GKP qubit as described in Eqs.~(\ref{gkpfinite0}) and (\ref{gkpfinite1}), we set the coefficients for Eq.~(\ref{ancilla}), $c_{t}$, to
\begin{equation}
c_{t}={e}^{-2\pi\kappa^2 (t-m)^2} \frac{\sqrt{2^{2t} (2 t)!}}{H_t(0)} , \label{coe}
\end{equation}
where ${H_t(x)}$ is the Hermite polynomial and ${H_t(0)}$ is determined so that the envelope of the output state becomes that of the approximated GKP qubit when the homodyne measurement outcome $x=0$~\cite{note2}.

In the second step, the squeezed light and the ancilla Fock state interact with each other by a Kerr medium with the strength of the nonlinearity $\chi$, as described in Fig.~\ref{fig2}(b)(ii) for the specific case $m = 2$. In this step, the squeezed light and the ancilla Fock state are entangled, where the phase of the squeezed light in phase space is rotated by phase $\theta_t=t\theta$ corresponding to Fock bases in Eq.~(\ref{ancilla}). 
After the interaction, the squeezed light and the ancilla Fock state evolve (up to normalization) as 
\begin{equation}
{e}^{i\hat{H}_{\rm CK}}\ket{\psi(\theta_0)}\ket{\phi}
 \propto \sum_{t=0}^{2m}c_{t}\ket{\psi(\theta_t)}\ket{2t}, 
\end{equation} 
where $\ket{\psi(\theta_t)}$ denotes the squeezed light with a phase parameter $\theta_t=t\theta$.
We consider the ancilla Fock state composed of an even number of Fock states.
This is because the probability of an odd number of Fock states becomes zero when the measurement outcome $x=0$ is postselected, where the probability of an odd number of Fock states for the position $x=0$, i.e., $|\braket{x=0|2t+1}|^2$, is zero. 
We mention that the amount of the phase rotation becomes $n$ times when we replace  $\ket{2t}$ with $\ket{2nt}$ as the Fock bases in Eq.~(\ref{ancilla}).

In step 3, the displacement operation, $D_{1}$, transforms the amplitude of the squeezed light by $\beta+i\gamma$ in phase space $( \gamma \in  \mathbb{R})$, as described in Fig.~\ref{fig2}(b)(iii) for the case $m$~=~2.
In this step, we set values $\gamma=2m\sqrt{\pi}$ so that the distance between squeezed light in the $q$ quadrature becomes $2\sqrt{\pi}$ which corresponds to the codewords for $\ket{\widetilde{0}}$, as shown in Fig.~\ref{fig2}(c) for the case $m = 2$. We note that $\gamma$ is set to $\gamma=m\sqrt{\pi}$ for the generation of  $\ket{\widetilde{+}}=(\ket{\widetilde{0}}+\ket{\widetilde{1}})\sqrt{2}$.
The displacement error in the $p$ quadrature, $\delta$, is defined by
\begin{equation}
\delta=\beta(1-{\rm cos}\theta)=\beta\left\{1-\sqrt{1-({\gamma}/{m\beta})^2}\right\}, \label{delta}
\end{equation}
which occurs the phase error on the generated GKP qubit for $m>1$. 
Thus, $\beta$ should be sufficiently large compared to $\gamma$ to reduce the phase error. 
Later we will see that we can ignore the displacement error $\delta$ in our scheme.

In step 4, the squeezed light and the ancilla Fock state interact with each other by the second Kerr medium with the strength of the nonlinearity $-\chi$, as shown in Fig.~\ref{fig2}(b)(iv). 
In this step, the rotation value for the $t$-th Fock state, $\ket{2t}$, in the second step should be turned into the negative value, $-t\theta$, by using the inverse strength of the interaction,$-\chi$. 
Then, all phases of the superposition of squeezed light become $\theta_t=0$ after the phase rotation by the interaction with the ancilla Fock state.
We note that each position of the squeezed lights in the $q$ quadrature corresponds to the target GKP qubit up to the displacement in the $p$ quadrature.

In step 5, the ancilla Fock state is measured by the homodyne detector in the $q$ quadrature, and the squeezed light is disentangled with the ancilla Fock state. 
The output state after the measurement of the ancilla Fock state depends on the measurement outcome $x$. 
Although the output state is closest to the target state $\ket {\widetilde{0}}_m $ in Eq.~(\ref{gkpfinite0}) when $x=0$, the probability of obtaining $x=0$ gets close to zero.
Thus, we implement the conditional measurement using the upper bound of the measurement outcome $v_{\rm up}$, where the generation succeeds when $|x|\leq v_{\rm up}$ so that the fidelity between the generated state and the target GKP qubit becomes a sufficient quality for FTQC.

In this step, when the homodyne measurement succeeds, we implement the displacement operation, $D_{2}$, on the squeezed light by $-\beta- m\delta/2$ in the $p$ quadrature, as shown in Fig.~\ref{fig2}(b)(v) for $m=2$. Otherwise, the GKP qubit generation fails (when $|x|> v_{\rm up}$).
Assuming the measurement outcome $x$, the generated state can be described by
\begin{equation}
\ket{\widetilde{0}}_{m,x,\delta} \propto   \sum_{t=-m}^{m} \int  c_{t+m} H_{t+m}(x) {e}^{-\frac{(s-2t\sqrt{\pi})^2}{2\Delta^2}} {e}^{-i\delta_t^m s}\ket{s}_q  ds, \label{gkpfinite3} 
\end{equation}
where we set the phase errors to $\delta_t^m= (m/2-|t|)\delta$.
The phase errors for the specific case $m=2$ are $(\delta_{-2}^2,\delta_{-1}^2,\delta_{0}^2,\delta_{1}^2,\delta_{2}^2)=(-\delta,0,\delta,0,-\delta)$, as shown in Fig.~\ref{fig2}(d).
We note that each of the coefficients of the generated state corresponds to the envelope's distribution of GKP qubit as described in Eq.~(\ref{coe}) with $x=0$, and the generated state corresponds to $\ket {\widetilde{0}}_m $ with $x=0$ and $\delta=0$ in Eq.~(\ref{gkpfinite0}). 
Thus, the proposed method generates the GKP qubit well by selecting appropriate parameters $m$, $\delta$, and $v_{\rm up}$.

\begin{figure}[b]
\centering\includegraphics[scale=1.0]{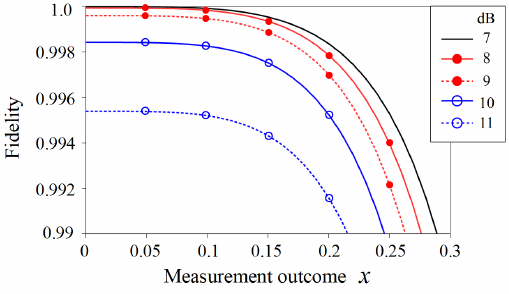}
\caption{Fidelity of  the generated GKP qubit as a function of the measurement outcome $x$ for the several squeezing levels, 7, 8, 9, 10, and 11 dB, where $x$ is obtained by the homodyne measurement of a superposition of Fock states in the $q$ quadrature. The displacement error in the $p$ quadrature, $\delta$, is assumed to be zero.}
\label{fig3}
\end{figure}

\section{Numerical calculations}\label{Sec4}
We consider the generation of the GKP qubit with the squeezed levels 7, 8, 9, 10, and 11 dB, where around 10 dB is sufficient to implement FTQC~\cite{fukui2018high,fukui2019high}.
In Fig.~\ref{fig3}, the fidelity of the generated GKP qubit for the target state is plotted as a function of the homodyne measurement outcome $x$ in the $q$ quadrature, assuming $m=2$ and $\delta=0$.
The fidelity, $F(x)$, is obtained by
\begin{equation}
F(x)=|\braket {{{\widetilde{0}}|\widetilde{0}}}_{m,x,\delta}|^2
\end{equation}
Numerical results show that our method with $m=2$ can generate the approximated GKP qubit with high fidelity, e.g., $F(x)\geq 0.99$ for $x\leq 0.2$.
In addition, our method is feasible in a wide range of the homodyne measurement, since the fidelity is stable for $|x|\leq 0.1$. This feasibility works to the advantage of the success probability of our scheme.

We note that the fidelity decreases when the squeezing level becomes larger than 10 dB.
This is because the number of peaks $m=2$ is not enough to approximate the envelope of the GKP qubit $\ket{\widetilde{0}}$, where the probability of the peaks with the amplitude $\pm 6\sqrt{\pi}$ can not be ignored with the large squeezing level.
To improve the fidelity for the generation of GKP qubit with the squeezing level larger than 10 dB, we may set the parameter for the number of peaks as $m=3$, i.e., the number of peaks of the generated state is 6. This is because the GKP qubit with a higher squeezing level has a larger number of peaks.
For $m=3$, the number of coefficients of the ancilla Fock state is set to 6 in Eq.~(\ref{ancilla}).
We found that the fidelities of the generated state with 10, 11, and 12 dB can be increased to 99.998, 99.986, and 99.938~\% by using the ancilla state with $m=3$, respectively. 
We note that the interaction strength $\chi$ does not change according to $m$.

\begin{figure}[t]
\centering\includegraphics[scale=1.0]{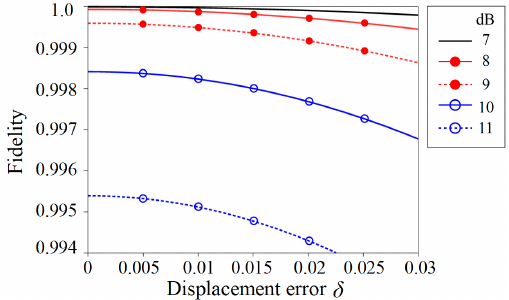}
\caption{Fidelity of  the generated GKP qubit with the homodyne measurement result $x=0$ as a function of the displacement error in the $p$ quadrature, $\delta$, for several squeezing levels, 7, 8, 9, 10, and 11 dB.}
\label{fig4}
\end{figure}

We then see the effect of the displacement error in the $p$ quadrature, $\delta$.
In Fig. \ref{fig4}, the fidelity of the generated GKP qubit with the homodyne measurement result $x=0$ is plotted as a function of the error $\delta$. 
Numerical results show that the effect of is $\delta$ expected to be ignored in the range of $\delta \leq 0.02$, where the fidelities decrease by only $\sim0.01$.
In particular, the condition of $\beta$ is obtained by $\beta(1-{\rm cos}\theta) \leq 0.02$ using Eq.~(\ref{delta}). 
In the case of $\gamma=4\sqrt{\pi}$, $\beta$ is needed to be more than $\sim314$ for $\delta \leq 0.02$. In general, this requirement for $\beta$ is much less than $\sim 4\times 10^{5}$, which is required to obtain an amount of a cross-Kerr interaction as described in Sec. II.
Thus, we ignore the displacement error $\delta$ in our scheme.

We finally calculate a mean fidelity as a function of the success probability.
Although the fidelity of the generated state with $x=0$ reaches the maximum value as shown in Fig.~\ref{fig3}, the success probability of our method becomes almost zero. 
Thus, we need to determine the adequate success probability for achieving the appropriate fidelity.
Here we introduce an upper bound $v_{\rm up}$ so that the GKP qubit generation succeeds when $|x|\leq v_{\rm up}$.
Then, the success probability with $v_{\rm up}$, $P_{\rm suc}(v_{\rm up})$, is calculated by
\begin{equation}
P_{\rm suc}(v_{\rm up})=\frac{1}{P_{\rm all}}\int_{-v_{\rm up}}^{v_{\rm up}}p(x)dx ,
\end{equation}
where $p(x)$ and $P_{\rm all}$ are defined as
\begin{eqnarray}
&p&(x)=\sum_{t=0}^{2m+1}|c_{2t} H_{2t}(x)|^2,\\
&P&_{\rm all}=\int_{-\infty}^{\infty}p(x) dx,
\end{eqnarray}
respectively.
In the calculation of the success probability of our scheme, we assume that the ancilla Fock states is prepared deterministically since we focus on the potential of the cross-Kerr interaction.
Then we define the mean fidelity, $\overline{F}(v_{\rm up})$, in order to verify the generated GKP qubit.
$\overline{F}(v_{\rm up})$ is calculated by
\begin{equation}
\overline{F}(v_{\rm up})= \int_{-v_{\rm up}}^{v_{\rm up}}F'(x)dx,
\end{equation}
where $F'(x)$ is the differential of $F(x)$.

\begin{figure}[t]
\centering\includegraphics[scale=1.0]{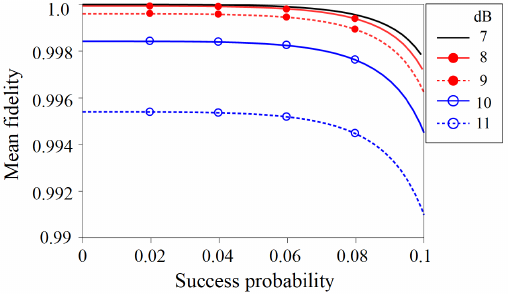}
\caption{Success probability and mean fidelity for the generated GKP states with the ancilla Fock state with $m$=2 for several squeezing levels, 7, 8, 9, 10, and 11 dB, assuming that the ancilla Fock states is prepared deterministically.}
\label{fig5}
\end{figure}

In Fig. \ref{fig5}, the mean fidelities between the generated state and the GKP qubit are plotted as a function of the success probability for the squeezing levels 7, 8, 9, 10, and 11 dB~ \cite{note3}.
In the numerical results for the ancilla Fock state with $m=2$, the mean fidelities are stable for the success probabilities $P_{\rm suc}(v_{\rm up})<0.06$ with $v_{\rm up}\sim0.15$.
Numerical results show that our method with $m=2$ can generate the GKP qubit with a feasible success probabilities, where the generated GKP qubit reaschs the fidelity larger than 99.9\%.

\begin{figure}[t]
\centering\includegraphics[scale=1.0]{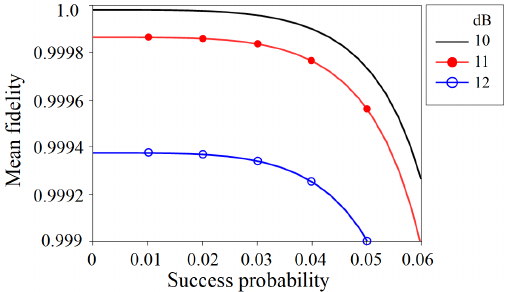}
\caption{Success probability and mean fidelity for the generated GKP states with the ancilla Fock state with $m$=3 for several squeezing levels, 10, 11, and 12 dB. The fidelities for $P_{\rm suc}(v_{\rm up} \to 0) \sim 0$ correspond to 99.998, 99.986, and 99.938~\% for the target GKP qubit with 10, 11, and 12 dB, respectively. }
\label{fig6}
\end{figure}

To perform FTQC with the GKP qubit, the fidelity for the GKP qubit with the squeezing level larger than 10 dB should be close to $\sim$1.
To improve the fidelity, we consider the number of peaks as $m=3$.
In Fig.~\ref{fig6}, we calculated the mean fidelities by using the ancilla Fock state with $m=3$.
The mean fidelities more than 99.9\% can be achieved for the success probabilities $P_{\rm suc}(v_{\rm up})=0.05$ with $v_{\rm up}\sim0.16$.
The fidelities for $P_{\rm suc}(v_{\rm up} \to 0) \sim 0$ correspond to 99.998, 99.986, and 99.938~\% for the target GKP qubit with 10, 11, and 12 dB, respectively.



\section{Discussion and conclusion}\label{Sec5}
In this paper, we have developed the method to generate the GKP qubit using the cross-Kerr effect between a squeezed light and a superposition of Fock states.
Our scheme overcomes the problems of the conventional method for the FTQC with the GKP qubits. 
We numerically show that our method has the potential to generate the GKP qubit with a high squeezing and a high fidelity. Thus our method could provide the potential way via the cross-Kerr interaction to realize FTQC with the GKP qubit.

We mention the experimental realization of the proposed method.
The preparation of a squeezed light, the achievable squeezing level of 15 dB has been reported~\cite{vahlbruch2016detection}, which is more than the required squeezing level for FTQC~\cite{fukui2018high,fukui2019high,noh2020fault,larsen2021fault,bourassa2021blueprint,tzitrin2021fault}. 
Regarding the cross-Kerr interaction, a small strength of the cross-Kerr interaction is enough to implement our scheme, e.g., $\theta \sim 10^{-5}$ which is achievable in the experiments~\cite{feizpour2015observation}.
In addition, the inverse of sign for the cross-Kerr interaction could be realized by using the measurement-induced interaction or Rubidium-based cross-Kerr medium~\cite{feizpour2015observation}.
Regarding the preparation of the ancilla Fock state, the maximum value of coefficients for the ancilla has been still three in an optical setup~\cite{yukawa2013generating}. 
Fortunately, there is a promising scheme to generate arbitrary coefficients of a superposition of Fock states in Refs.~\cite{su2019conversion,tzitrin2020progress}, and the experimental realization for the photon number resolving detector, a key role in Refs.~\cite{su2019conversion,tzitrin2020progress}, has been demonstrated~\cite{lita2008counting,fukuda2011titanium,endo2021quantum}.
Additionally, in a realistic experimental setup, there are the effects derived from a photon loss during the preparation of the ancilla Fock state, the imperfection of the cross-Kerr interaction, and the inaccuracy of the homodyne measurement. These effects may degrade the fidelity of the generated GKP qubit. We will investigate the effect of a photon loss in a future work since it is beyond the scope of this paper.
Nevertheless, our method will provide an efficient way to generate the GKP qubit required for FTQC, once appropriate technologies for the preparation of the ancilla Fock state and the cross-Kerr interaction are available in an experimentally feasible way.

We also mention the possibility to obtain the cross-Kerr interaction by using the decomposition technique~\cite{sefi2011decompose,sefi2013measurement,kalajdzievski2019exact} and the multi-mode nonlinear coupling~\cite{sefi2019deterministic}.
For the decomposition technique, the quantum gate for the $n$-order Hamiltonian with $n>3$ can be approximated by the sequential gates consisting of less than the $n'$-order Hamiltonians with $n'<n$. 
Thus, our proposed scheme can be implemented in an optical setup, by replacing the cross-Kerr interaction with the cubic phase gate and Gaussian operations. 
In an optical setup, the cubic phase gate can be implemented by the measurement-induced nonlinear interaction~\cite{gottesman2001encoding,filip2005measurement,sefi2013measurement,sabapathy2018states}, where the cubic phase state used for an ancilla in the nonlinear interaction has been demonstrated experimentally~\cite{yukawa2013generating}. 
For the multi-mode nonlinear coupling, Ref.~\cite{sefi2019deterministic} introduced the multi-mode gate for modes 1 and 2, which is composed of only $\hat{q}$ operators such as $e^{i \hat{q_1}^n \hat{q_2}^{n'}}$ with $n+n'>3$.
Although the cross-Kerr gate is composed of $\hat{q}$ and  $\hat{p}$ operators, which is described as $e^{i( \hat{q_1}^2+\hat{p_1}^2)(\hat{q_2}^{2}+\hat{p_2}^{2})}$, we may obtain the cross-Kerr interaction by applying multi-mode nonlinear coupling to the cross-Kerr gate.
Lastly, it will be worth comparing our scheme with the breeding protocol with a phase estimation~\cite{terhal2016encoding,weigand2018generating}, which is a promising way to generate the GKP qubit.
The difficulty of our scheme is the requirement of the ancilla of a superposition of Fock states, and a Kerr medium, while the difficulty of the breeding protocol is the requirement of the ancilla of the squeezed cat state in an optical setup.

In terms of fidelity and the success probability, our scheme could be comparative to the breeding protocol in an optical setup~\cite{weigand2018generating} when the ancilla Fock state for our method and squeezed cat state for the breeding protocol can be prepared deterministically.

\acknowledgments
This work was partly supported by JST [Moonshot R$\&$D][Grant No. JPMJMS2064], JST [Moonshot R$\&$D][Grant No. JPMJMS2061], JSPS KAKENHI [Grant No. 20K15187], UTokyo Foundation, and donations from Nichia Corporation. 
M.E. acknowledges supports from Research Foundation for Opto-Science and Technology.
\bibliography{ref.bib}

\begin{thebibliography}{91}
\bibitem [{not({\natexlab{b}})}]{note4}%
  \BibitemOpen
  \href@noop {} {}\bibinfo {note} {{The GKP qubit generated in an ion trap and
  superconducting circuit quantum electrodynamics could be more suitable for
  gate-based quantum computation towards fault-tolerant quantum
  computation}}\BibitemShut {NoStop}%
\bibitem [{not({\natexlab{c}})}]{note6}%
  \BibitemOpen
  \href@noop {} {}\bibinfo {note} {{We should mention that there is room to
  find the method to reduce the probability of the misidentifying the bit value
  of the state in Ref.~\cite{pirandola2004constructing} and to improve the
  threshold for FTQC with the GKP qubit}}\BibitemShut {NoStop}%
\bibitem [{not({\natexlab{e}})}]{note2}%
  \BibitemOpen
  \href@noop {} {}\bibinfo {note} {{ In this work, we have not investigated the
  capabilities of the device for the preparation of ancilla Fock state with
  $m$. {R}ef.\cite{su2019conversion} shows that the device with $l$ modes of
  Gaussian states can prepare the arbitrary coefficients up to $\ket{k}$ with
  $k=(l+2)(l-1)/2-1$, while it is open problem that the number of $l$ and the
  probability of preparation of ancilla Fock state with $m$ are
  required}}\BibitemShut {NoStop}%
\bibitem [{not({\natexlab{f}})}]{note3}%
  \BibitemOpen
  \href@noop {} {}\bibinfo {note} {{ Although there is no chance for getting
  the homodyeme measurement outcome $x=0$, we have calculated the mean fidelity
  ${F}(v_{\rm up})$ as the fidelity $F(x=0)$ in Eq. (19) for the sake of
  convenience}}\BibitemShut {NoStop}%
\end{thebibliography}
\end{document}